\documentclass[twocolumn,preprintnumbers,amsmath,amssymb]{revtex4-1}

\usepackage{graphicx}
\usepackage{dcolumn}
\usepackage{bm,epsfig}
\usepackage{epstopdf}

\begin{document}

\title{Filling the holes in the CaFe$_{4}$As$_{3}$ structure: synthesis and magnetism of CaCo$_{5}$As$_{3}$}

\author{P. F. S. Rosa$^{1}$, B. L. Scott$^{1}$, F. Ronning$^{1}$, E. D. Bauer$^{2}$, and J. D. Thompson$^{1}$}

\affiliation{
$^{1}$ Los Alamos National Laboratory, Los Alamos, New Mexico 87545, U.S.A.}
\date{\today}

\begin{abstract}
Here we investigate single crystals of CaCo$_{5}$As$_{3}$  by means of single crystal X-ray diffraction, microprobe, 
magnetic susceptibility, heat capacity, and 
pressure-dependent transport measurements. CaCo$_{5}$As$_{3}$ shares the same structure of CaFe$_{4}$As$_{3}$ with an additional Co atom filling a lattice vacancy
 and undergoes a magnetic 
transition at $T_{M} = 16$~K associated with a frustrated magnetic order. CaCo$_{5}$As$_{3}$ displays metallic behavior 
and its Sommerfeld coefficient ($\gamma = 70$~mJ/mol.K$^{2}$) indicates a moderate enhancement of 
electron-electron correlations. Transport data under pressures to $2.5$~GPa reveal a suppression of $T_{M}$ at a rate of $-0.008$~K/GPa. First-principle electronic structure calculations show a complex 3D band structure 
and magnetic moments that depend on the local environment at each Co 
site. Our results are compared with previous data on CaFe$_{4}$As$_{3}$ and provide a scenario for a magnetically frustrated ground state in this family of compounds.

 \end{abstract}

\maketitle

\section{INTRODUCTION}

Transition-metal pnictides and chalcogenides have attracted renewed interest in the past decade. From the discovery of high-temperature superconductivity in the Fe-based superconductors \cite{Kamihara,Rotter} to the 
experimental realization of a Weyl semimetal state in TaAs \cite{TaAs1,TaAs2,TaAs3}, these materials have become a promising avenue for the investigation of new states of matter. Although distinct phenomena emerge in different families 
of transition-metal pnictide compounds, a strong correlation between crystal structure and physical properties is commonplace. In the case of TaAs, spin-orbit coupling and inversion symmetry breaking are 
key to realizing a Weyl semimetal state. In the case of FeAs-based superconductors, the layered tetragonal structure containing planes of FeAs$_{4}$ tetrahedra is present in all superconducting families.

The search for new materials not only furthers understanding of the relationship between structure and materials' properties but also can lead to the discovery of new states that emerge from novel
structural arrangements. For instance, CaFe$_{4}$As$_{3}$ was synthesized soon after the report of superconductivity in Ba$_{1-x}$K$_{x}$Fe$_{2}$As$_{2}$ at $T_{c} = 38$~K \cite{CaFeAs1,CaFeAs2}. 
CaFe$_{4}$As$_{3}$ crystallizes in an orthorhombic structure with interpenetrating FeAs strips that also contain FeAs$_{4}$ tetrahedra. As shown in Fig.~1a, these strips have 
finite width in the $ac$-plane and are connected through fivefold coordinated Fe atoms [Fe(4)], creating channels that host Ca atoms. 
Thermodynamic measurements on CaFe$_{4}$As$_{3}$ (`143') single crystals reveal a
 second order transition at $T_{N}=89.6$~K. Neutron diffraction measurements find that the transition at $T_{N}$ is to a longitudinal ($||~b$) incommensurate (IC) spin-density-wave (SDW) order with a $3$D
Heisenberg-like 
critical exponent \cite{Neutrons}. At $T_{2}=25.6$~K, there is a first order transition to a commensurate order due to the loss of a degree of freedom associated with a soft mode present in the IC state. 
Band-structure calculations in the nonmagnetic regime appear to be consistent with a nesting instability associated with the planar Fermi sheets from cross-linked FeAs strips \cite{Neutrons}. 
Subsequent first-principle calculations, however, note that the nesting concept would be only valid in the linear perturbation regime over the calculated nonmagnetic state,
 and, as a consequence, inapplicable for a strongly magnetic state.
In fact, the ordered moments extracted from neutron diffraction below $T_{2}$ are
$2.2$~$\mu_{B}$ and $2.4$~$\mu_{B}$ per Fe$^{2+}$ (fourfold) and Fe$^{+}$ (fivefold) sites, respectively. These 
 values are considerably larger than those in Fe-based superconductors ($0.4-1$ 
$\mu_{B}/$Fe), but comparable to the moment of strongly correlated FeTe ($2.1$~$\mu_{B}$) \cite{Haule}.  Moreover, the magnetic 
structure realized in CaFe$_{4}$As$_{3}$ is similar to that of Fe$_{1.14}$Te \cite{Fe114Te}.

Theoretical and experimental reports on the Fe$Pn$-based materials ($Pn =$ pnictogen, chalcogen) show that the Fe moment is sensitive 
to the Fe-$Pn$ hybridization and tends to increase with increasing 
Fe-$Pn$ distance \cite{Haule, Neutrons2, Granado}. In fact, the Fe-Te distance in Fe$_{1.14}$Te ($2.59$~\AA) is comparable to the Fe-As distances in 
CaFe$_{4}$As$_{3}$ ($2.4-2.61$~\AA) but significantly larger than corresponding distances
in the small-moment parent compounds CaFe$_{2}$As$_{2}$ ($2.37$~\AA) and BaFe$_{2}$As$_{2}$ ($2.39$~\AA). Density functional calculations suggest that thicker Fe$Pn$ planes stabilize magnetic structures
found in FeTe and CaFe$_{4}$As$_{3}$  due to the effect of bond angles on second and third nearest-neighbor interactions \cite{Moon}.  Further, as noted in Ref.~\cite{Neutrons}, the sequence of phase transitions 
observed in CaFe$_{4}$As$_{3}$ also 
occurs in TbMnO$_{3}$ \cite{TbMnO3} and Ni$_{3}$V$_{2}$O$_{8}$ \cite{Ni3V2O8}, and can be explained by a combination of competing exchange interactions 
and easy-axis anisotropy. Finally, the simplest of the FeAs-based compounds, FeAs, is highly orthorhombic and also orders in an
IC-SDW. First-principle calculations on FeAs show that there is no correspondence whatsoever between
the computed real part of the susceptibility and the ordering vector \cite{Mazin}. This result indicates that the 
common ordering pattern is not nesting-driven in these materials. In particular, next-nearest-neighbors in FeAs have parallel spins and the Fe-As-Fe bond
angle is close to 90$^{o}$ whereas nearest neighbors have opposite spins and much flatter angles, consistent with the Goodenough-Kanamari rules. Though these rules 
were derived for dielectrics, the observed agreement hints to the presence of local moment physics.

All of the above suggests a scenario in which magnetic frustration plays a role and IC-SDW order in CaFe$_{4}$As$_{3}$ arises from competing magnetic
couplings. Geometrical frustration in metals and insulators has been explored extensively in pyrochlore, Kagom\'{e}, and triangular lattices \cite{Pyro,Triang} but magnetic frustration also may emerge in 
the absence of geometrical frustration due solely to competing long-ranged exchange interactions. Experimental examples in tetragonal systems have been identified recently in Ce-based metals 
with non-collinear magnetic structures, such as 
CeRhIn$_{5}$ and CeAgBi$_{2}$ \cite{David,Thomas}. A complete understanding of exchange-driven magnetic frustration in transition metal-based systems without geometrical frustration is, however, still elusive and the question 
of whether there is a universal way of modeling it in itinerant systems remains unanswered. Magnetic frustration has been proposed to be an important ingredient to describe Fe-based materials 
\cite{Si} and, 
therefore, the `143' orthorhombic structure and its derivatives may be a good platform to address this question.

Chemical substitution studies together with the investigation of similar parent compounds often provide important insights on the nature of the magnetism in structurally related systems. 
Previous studies on Ca(Fe$_{1-x}$Co$_{x}$)$_{4}$As$_{3}$ ($0 \leq x \leq 0.32$) grown in Sn flux showed a slow suppression of the IC-SDW state with $x$, whereas the IC-C transition rapidly disappeared for $x> 0.1$ \cite{CaFeAs3}. 
Single crystals with $x > 0.32$ could not be synthesized, which was suggested to be caused by the solubility
limit for the `143'  structure. We note, however, that the presence of a stable binary compound such as CoSn$_{2}$ could be detrimental to the formation of the Co `143' analog, and other fluxes may solve this issue.

Here we report the synthesis of CaCo$_{5}$As$_{3}$ using Indium flux. CaCo$_{5}$As$_{3}$ shares the 
same orthorhombic structure of CaFe$_{4}$As$_{3}$ but has an additional transition metal site filling the `holes' 
between Ca atoms. This new Co-based compound orders below $T_{M} = 16$~K in a magnetic 
state that cannot be ascribed to a simple magnetic structure.
CaCo$_{5}$As$_{3}$ is a good metal at low temperatures and applied pressures to $2.5$~GPa suppress $T_{N}$ at a rate of only $-0.008$~K/GPa. Electron-electron correlations are modest, as evidenced by a Sommerfeld coefficient ($\gamma$) of $70$~mJ/mol$_{f.u.}$.K$^{2}$. First-principle calculations show a complex three-dimensional band structure in which the magnitude of the magnetic moments depends on the local environment at each Co 
site. Our results support a scenario in which frustration, driven by competing exchange interactions, play a key role.


Single crystalline samples of CaCo$_{5}$As$_{3}$ were grown by the In-flux technique. The mixture of elements was placed in an alumina crucible and sealed in a quartz tube under vacuum. The sealed tube was heated to 
1000$^{\circ}$C for 12 h and then cooled to 400$^{\circ}$C at 6$^{\circ}$C/h. The flux was then removed by centrifugation. The crystallographic structure was verified by single-crystal diffraction at room 
temperature. In addition, several samples were characterized by elemental analysis using a commercial Energy Dispersive Spectroscopy (EDS) microprobe. 

A Quantum Design superconducting quantum interference device and small-mass calorimeter that employs a quasi-adiabatic thermal relaxation technique  were used to measure magnetic and specific heat properties of the crystals. The electrical resistivity was obtained using a low-frequency ac resistance bridge and a four-contact configuration. Needle-like crystals were 
mounted in a hybrid piston-cylinder pressure cell with Daphne oil 7373 as the pressure medium. The change in $T_{c}$ of a piece of Pb served as a manometer.
The electronic structure was calculated using density functional theory with an all electron full potential linearized augmented 
plane wave 
method with local orbitals. Calculation were implemented in the code WIEN2k using a generalized gradient approximation \cite{Blaha, Perdew}. Spin orbit coupling was included for computing the density of states.

\section{RESULTS}

\begin{figure}[!ht]
\begin{center}
\hspace{-0.3cm}
\includegraphics[width=1.0\columnwidth]{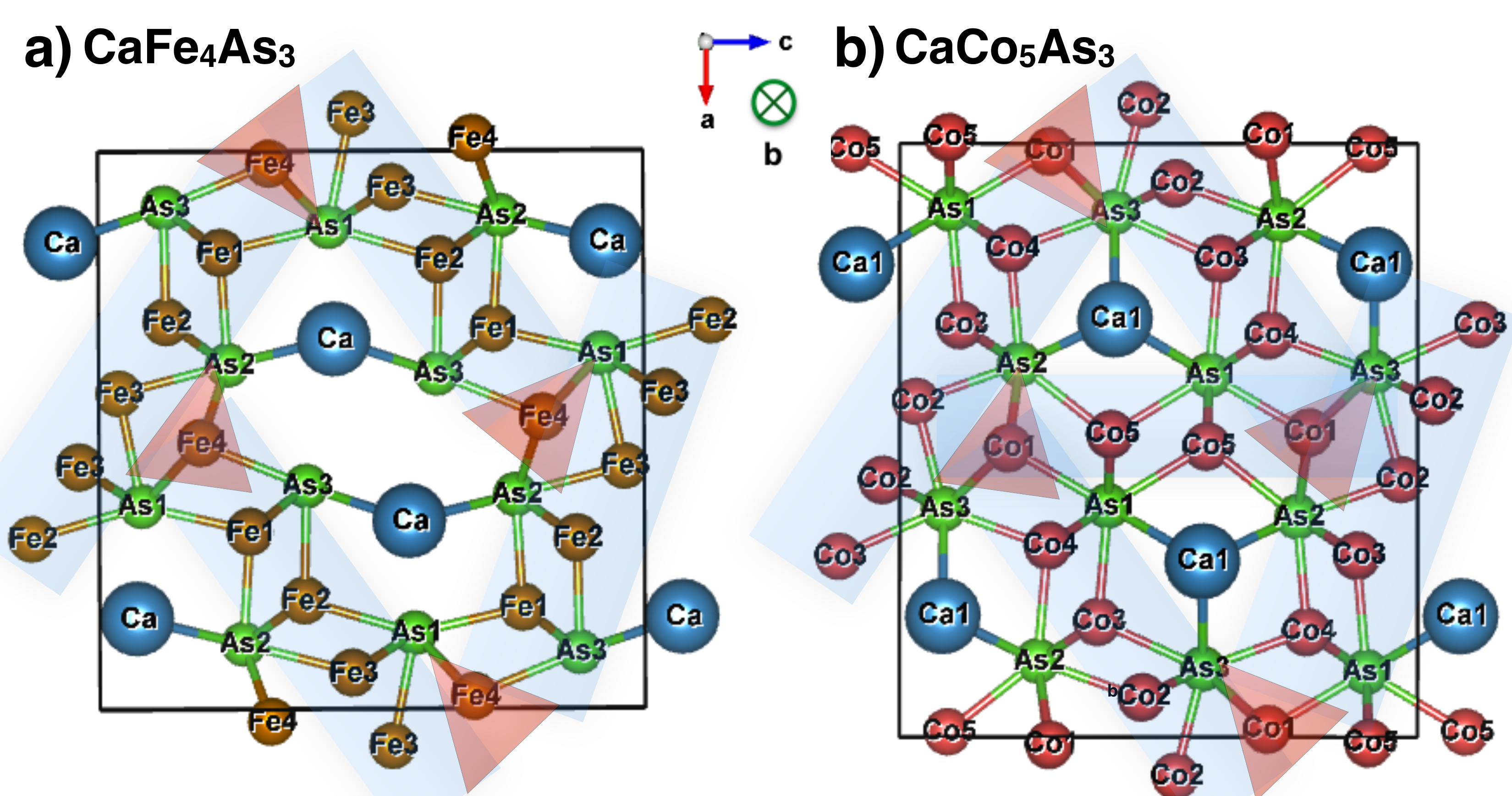}
\vspace{-0.75cm}
\end{center}
\caption{Crystal structures of (a) CaFe$_{4}$As$_{3}$ (Ref. \cite{CaFeAs1}) and (b) CaCo$_{5}$As$_{3}$ (this work). The elements Ca, Co, Fe, and As are represented by blue, red, gold, and green spheres.} 
\label{fig:Fig1}
\end{figure}

Figure~\ref{fig:Fig1} shows the orthorhombic crystal structure shared by CaFe$_{4}$As$_{3}$ \cite{CaFeAs1}
and CaCo$_{5}$As$_{3}$ (this work). 
 In CaFe$_{4}$As$_{3}$, FeAs$_{4}$ tetrahedra form infinitely long segments along the $b$ direction. In the $ac$ plane, FeAs$_{4}$
 tetrahedra strips containing structurally inequivalent Fe$^{2+}$ ions [Fe(1), Fe(2), and Fe(3) sites in the blue region] are connected by fivefold 
 Fe$^{+1}$ ions [Fe(4) site] to form a rectangular network that encloses two Ca atoms.  Although the same crystal structure is realized in
 CaCo$_{5}$As$_{3}$, the ``hole" between Ca atoms is filled by CoAs$_{4}$ tetrahedra at the Co(5) site, resulting in a more three-dimensional network with five inequivalent Co sites. 
We note that CaFe$_{5}$As$_{3}$ can be obtained via high-pressure synthesis and crystallizes in a monoclinic ($P2_{1}/m$) structure \cite{CaFe5As3}.
 Physical properties of CaFe$_{5}$As$_{3}$ are unknown, but it is noteworthy that the Fe- and Co-`153' compounds form in different structure types; whereas, the Fe `143' and Co `153' materials crystallize in the same space group.

 Table I summarizes the structural parameters of  CaCo$_{5}$As$_{3}$ determined by single crystal X-ray diffraction. The lattice parameters are close to those of CaFe$_{4}$As$_{3}$ 
($a = 11.852(3)$ \AA, $b = 3.7352(6)$ \AA, $c = 11.5490(18)$ \AA) 
 and the volume of the unit cell increases by 0.5 \%, likely due to the additional transition metal site.

 \begin{table}[h]

\caption{Crystallographic data of CaCo$_{5}$As$_{3}$ determined by single crystal X-ray diffraction. }\label{Table-crystal1}
\begin{tabular*}{1\linewidth}{@{\extracolsep{\fill}}ll}
\hline
\hline
   Crystal system            &                Orthorhombic	$\hspace{10.0cm}$                                                                                                    \\
   Space group               &         	   Pnma (62)	                                                                                                \\
   Temperature               &                296(2) K                                                                                                                            \\
   Wavelength                &                0.71073 \AA                                                                                                       \\                                                                                                          
   2$\theta_{min}$           &                2.49$^{o}$                                                                                                         \\
   2$\theta_{max}$           &                28.43$^{o}$                                                                                                       \\
   Index ranges              &                -16 $\leqslant$ h $\leqslant$ 16,    -4 $\leqslant$ k $\leqslant$ 5, -14 $\leqslant$ l $\leqslant$ 14              \\
\end{tabular*}
\begin{tabular*}{1\linewidth}{@{\extracolsep{\fill}}lc}
\hline
\hline						
Formula weight                          &	559.51 	       	       \\
  a (\AA)\                               &	12.425(2) 	           \\
    b (\AA)\                               &	3.8101(6) 	           \\
  c (\AA)\	                            &	10.8513(17) 	        \\
  Volume (\AA$^{3}$)	                &	513.71(14) 	     \\
  $\mu$ (Mo K$\alpha$) (cm$^{-1}$)      &    16.348             \\
  Goodness-of-fit on F$^{2}$            &	1.019  	                \\
  $R(F)$ for $F_{o}^{2}$ $>$   2$\sigma (F_{o}^{2})^{a}$            &     0.0222                                  \\
  $R_{w}(F_{o}^{2})^{b}$                &   0.0495                  \\
\hline
\hline
\end{tabular*}
\begin{tabular*}{1\linewidth}{@{\extracolsep{\fill}}ll}
 $^{a}R(F)$ = $\sum\mid\mid F_{o}\mid -\mid F_{c}\mid\mid/\sum\mid F{_o}\mid$ & \\
   $^{b}R_{w}(F_{o}^{2})$	= $[\sum[w(F_{o}^{2}-F_{c}^{2})^{2}]/\sum wF_{o}^{4}]^{1/2}$ &   \\ 	 \end{tabular*}				
\end{table}

 Table II displays the atomic coordinates and equivalent
 displacement parameters for CaCo$_{5}$As$_{3}$. The site symmetries are identical to those found in CaFe$_{4}$As$_{3}$ and the atomic coordinates are also very similar.
 The refined occupancy of CaCo$_{5}$As$_{3}$ is 100 \% for all sites, in agreement with the microprobe result CaCo$_{4.9(1)}$As$_{3.1(1)}$. 

\begin{table}[h]
\caption{Atomic coordinates and equivalent displacement parameters (\AA$^{2}$) for CaCo$_{5}$As$_{3}$ determined by single crystal x-ray diffraction at room temperature. U$_{eq}$ is defined as one third of the trace of the orthogonalized U$^{ij}$ tensor.}\label{Table-Co}
\begin{center}
\begin{tabular}{l@{\hspace{0.3cm}}c@{\hspace{0.2cm}}c@{\hspace{0.2cm}}c@{\hspace{0.6cm}}c@{\hspace{0.6cm}}c@{\hspace{0.6cm}}c}								
		\hline													
        Atom           & Wyck.      & Occ.   &	  x	          &	      y	         &	  z	        &   U$_{eq}$    \\
		\hline													
         As(1)	       &	4c	    &	1	&	0.11107	          &	0.25000	             &	0.09824	        &	0.005	\\
         As(2)	       &	4c	    &	1	&	0.12911	          &	0.25000	             &	0.73262 		&	0.005	\\
         As(3)	       &	4c	    &	1	&	0.38532	          &	0.25000	             &	0.91474 		&	0.005	\\
         Co(1)	       &	4c	    &	1	&	0.01244	          &	0.25000	             &	0.29001 		&	0.008	\\
         Co(2)	       &	4c	    &	1	&	0.06633             &   0.25000   		   & 0.53238     		&	0.005	\\
         Co(3)	       &	4c	    &	1	&	0.30513   		& 0.25000			   &    0.11698		 & 0.005	\\
         Co(4)	       &	4c	    &	1	&	0.32191		&    0.25000 		&   0.71380  		 &	0.005	\\
         Co(5)	       &	4c	    &	1	&	0.00960    	&  0.25000      		&  0.90666 		&	0.006	\\
         Ca(1)	       &	4c	    &	1	&	0.29548   		& 0.25000  		&  0.41600 		&	0.006	\\
 					
	  \hline      					
	    \hline
\end{tabular}
  \end{center}
\end{table}

\begin{figure}
\begin{center}
\includegraphics[width=0.75\columnwidth,keepaspectratio]{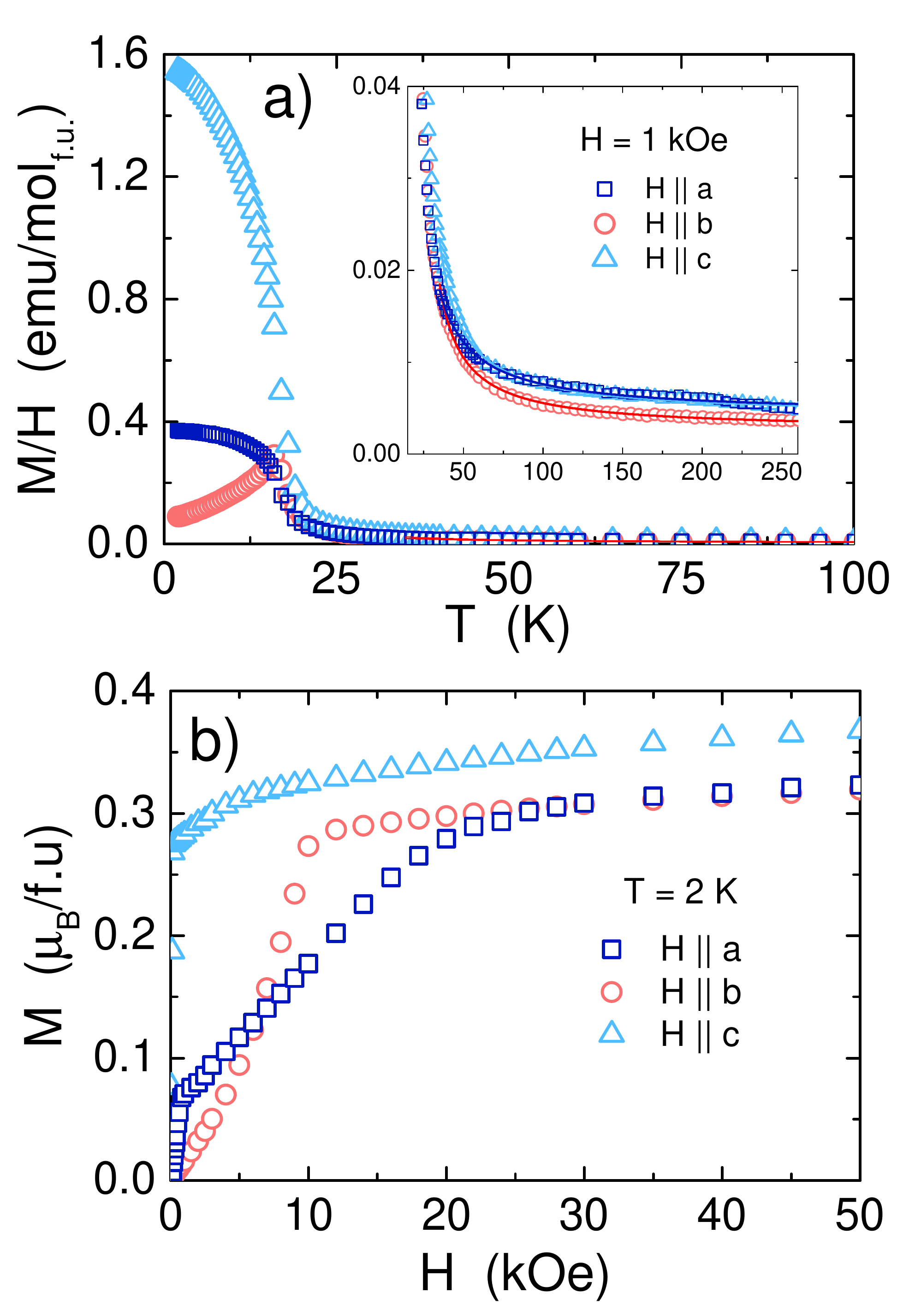}
\vspace{-0.75cm}
\end{center}
\caption{a) Temperature dependence of the magnetic response, $M(T)/H$, of CaCo$_{5}$As$_{3}$ in a field of $1$~kOe applied along the $a$-axis, $b$-axis, and $c$-axis.
Inset shows the fits of the high-temperature data to a modified Curie-Weiss law. 
b) Anisotropic field dependence of the magnetization at $2$~K.}
\label{fig:Fig2}
\end{figure}

Figure~\ref{fig:Fig2}a shows the temperature-dependent magnetic response, $M(T)/H$, to a field of $1$~kOe applied parallel to the a-axis (dark blue squares), b-axis (red circles) and c-axis (light blue triangles). These 
results are from zero-field cooled samples. A clear peak at $16$~K for $H~||~b$ is characteristic of the onset of antiferromagnetic order. For $H~||~a$ and $H~||~c$, however, $M(T)/H$ is typical of 
ferromagnets at the same transition temperature. This clear difference suggests the presence of anisotropic magnetic interactions between Co spins with ferromagnetic and antiferromagnetic components.

Above the ordering temperature where $M$ is linear in $H$, $M/H \equiv \chi$, the uniform susceptibility. For all directions, $\chi(T)$ can be fit to a Curie-Weiss (CW) law plus a $T$-independent Pauli term, 
i.e., $\chi(T)=\chi_{0} + C/(T - \theta)$, in the range from $35$~K to $350$~K. The extracted values of 
$\chi_{0}$  are $0.004(2)$~emu/mol$_{\mathrm{f.u.}}$ and $0.00257(3)$~emu/mol$_{\mathrm{f.u.}}$ for $H~||~a,c$ and $H~||~b$, respectively. 
The effective moment from these fits also is weakly anisotropic, 
 $1.48(3)$ $\mu_{B}/$f.u. ($H~||~a,c$) and  $1.38(1)$ $\mu_{B}/$f.u. ($H~||~b$) both of which are smaller than the theoretical value expected for Co$^{2+}$ ($3.87$~$\mu_{B}$ using $J=S$).
  Naively, the effective moment per Co would be $0.26$~$\mu_{B}$ (from one-fifth of a polycrystalline average of $\mu_{\mathrm{eff}}$) but, as we will come to later, the local
 environment and the hybridization between Co and As atoms are different for each Co site.
From the CW fits, the Curie-Weiss temperatures are
 $\theta = 18(4)$~K and $20(2)$~K for $H~||~a,c$and $H~||~b$, respectively. These temperatures are identical within experimental error, positive, which is characteristic of dominant ferromagnetic interactions, and close in value to $T_{N}$, typically an indication of little, if any, frustration.

From these parameters that describe the uniform susceptibility, it would be reasonable to expect that CaCo$_{5}$As$_{3}$ could be a simple (Stoner-like) ferromagnet. As evidenced from the sharp peak in $\chi(T)$ for $H~||~b$, 
this simple expectation obviously is incorrect, and the magnetic structure below $16$~K must be more complex. A non-trivial magnetic structure is further reflected in the magnetization as a function of 
magnetic field, shown in Fig.~\ref{fig:Fig2}b. For $H~||~a$, the magnetization initially increases quickly as expected for a soft ferromagnetic, but instead of continuing smoothly to saturation, there is a clear anomaly
 in $M_{a}(H)$ at $\sim 0.8$~kOe, pointing to a change in magnetic structure. At higher fields, $M_{a}(H)$ smoothly approaches a regime in which it has a small but finite slope 
 $dM/dH=0.0053$~emu/mol$_{\mathrm{f.u.}}$ that is about $1.5$ times larger than $\chi_{0}$ for this orientation. We note that $M_{c}(H)$ has a similar high-field finite slope. 
When $H~||~b$, $M_{b}(H)$ increases monotonically at low fields and crosses $M_{a}$. At $10$~kOe, there is a kink in $M_{b}(H)$, possibly indicating a change in magnetic structure
 above which $dM/dH$ approaches a constant value of $0.0063$ emu/mol$_{\mathrm{f.u.}}$ that is about 2.5 times larger than $\chi_{0}$.
These values of $dM/dH$ above $35$~kOe suggest either that the conduction band density of states at the Fermi energy ($N_{E_{F}}$) has been Zeeman split (by about $0.2$~meV) to give a $N_{E_{F}}$ that is 
1.5-2.5 times larger 
than at low fields or that there may be one or more additional changes in magnetic structure at fields above 50 kOe. From magnetization and susceptibility measurements, the most likely scenario is 
that CaCo$_{5}$As$_{3}$ adopts a non-trivial ferrimagnetic order in zero-applied field and that its structure is susceptible to field-induced changes. We will return to this possibility, but clearly,  
high-field measurements as well as microscopic measurements (e.g. neutron diffraction and nuclear magnetic resonance)
would be valuable to solve the magnetic structure of CaCo$_{5}$As$_{3}$.

\begin{figure}[!ht]
\begin{center}
\includegraphics[width=0.75\columnwidth,keepaspectratio]{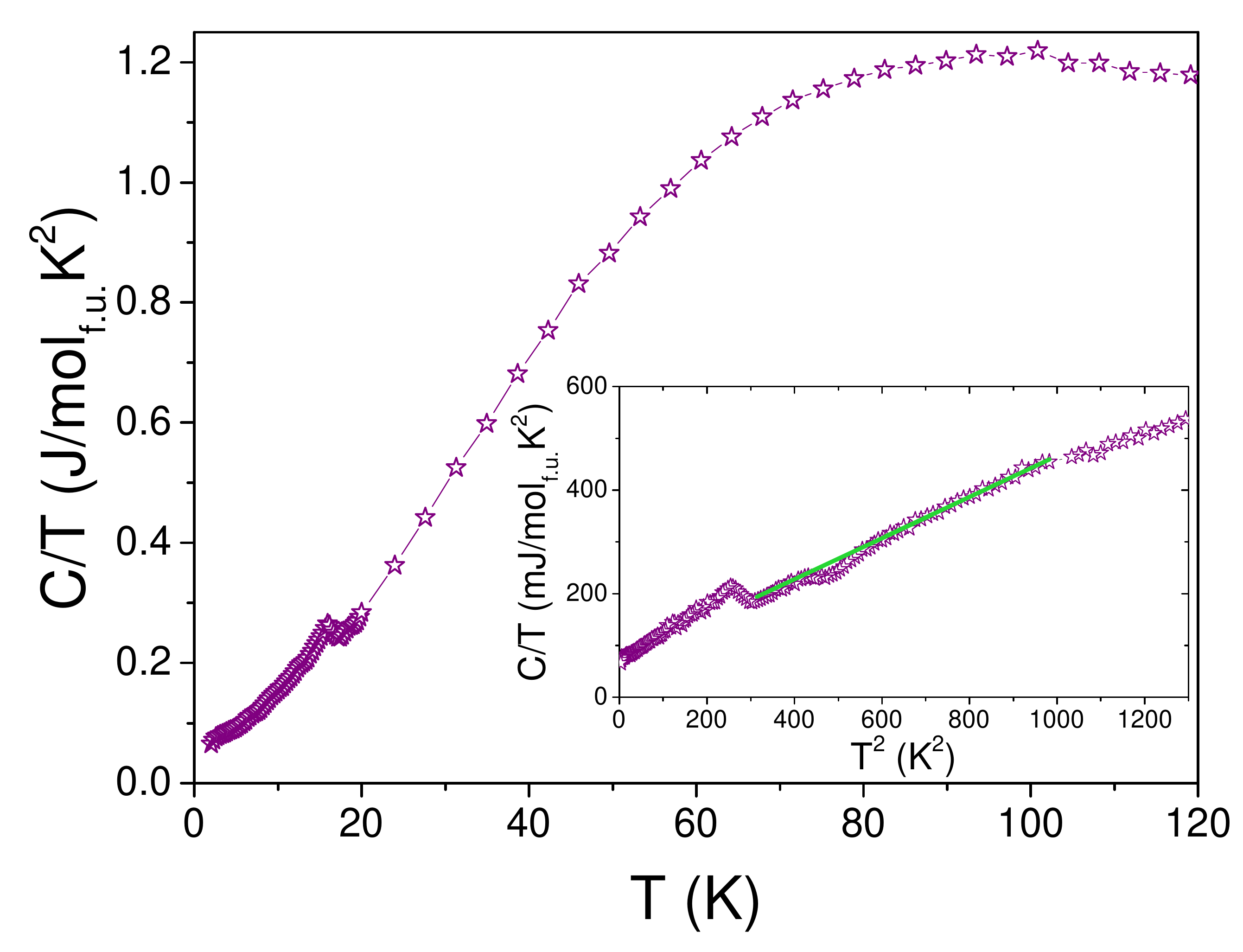}
\vspace{-0.7cm}
\end{center}
\caption{a) Specific heat data as a function of temperature. Inset shows a linear fit (solid line) 
in a $C/T$ vs $T^{2}$ plot.}
\label{fig:Fig3}
\end{figure}

Specific heat data of CaCo$_{5}$As$_{3}$ also display a transition at $16$~K (Figure~\ref{fig:Fig3}). Above the transition, a fit of the data to
$C/T=\gamma+\beta T^{2}$ yields an electronic coefficient of $\gamma = 70$~mJ/mol$_{\mathrm{f.u.}}.$K$^{2}$ (inset of Fig. 3). This value of $\gamma$ should be 
taken with caution because of the rather small window (15~K) over which the fit was made. Nevertheless, the same fit 
 below $T_{N}$ also yields 
$\gamma~=~70$~mJ/mol$_{\mathrm{f.u.}}.$K$^{2}$, in surprising agreement with the high-$T$ value. From these fits above and below $T_{N}$, entropy 
obviously is not conserved, suggesting that there may be another phase transition below the lowest temperature ($2$~K) of these measurements. 
Within the free-electron model, the Pauli susceptibility is defined as $\chi_{p}=\mu_{B}^{2}N(E_{F})$ where $N(E_{F})$ can be obtained 
via $\gamma = (1/3)\pi^{2}k_{B}^{2}N(E_{F})$. Here $\mu_{B}$ is the Bohr magneton constant and $k_{B}$ is the Boltzmann constant.
Using the high-$T$ experimental value of $\gamma$, the calculated Pauli susceptibility is $\chi_{p}= 1.3 \times 10^{-3}$ emu/f.u. which is smaller than
the values of $\chi_{0}$ obtained from the magnetic susceptibility ($2.6 \times 10^{-3}$ and $3.7 \times 10^{-3}$ emu/f.u. for $H~||~b$ and $H\perp b$, respectively).
This simple comparison between $\chi_{p}$ and $\chi_{0}$ implies that $\chi_{0}/\gamma > 1$ and suggests that the conductions electrons have FM correlations above $T_{N}$.

As shown in Fig.~4a, the temperature dependence of the electrical resistivity, $\rho(T)$, of 
CaCo$_{5}$As$_{3}$ along the $b$-axis displays metallic behavior and a curvature that is likely due to scattering of conduction electrons by phonons and $s$-$d$ interband scattering.
The low-$T$ data display a clear kink at $T_{N}$, followed by a decrease due to the decrease in spin disorder scattering.  The resistance ratio $\rho_{300 K}/\rho_{2 K} \sim 25$ is five times larger than that in the Fe `143' counterpart and indicates good crystallinity.

\begin{figure}[!ht]
\begin{center}
\hspace{-0.75cm}
\includegraphics[width=0.75\columnwidth,keepaspectratio]{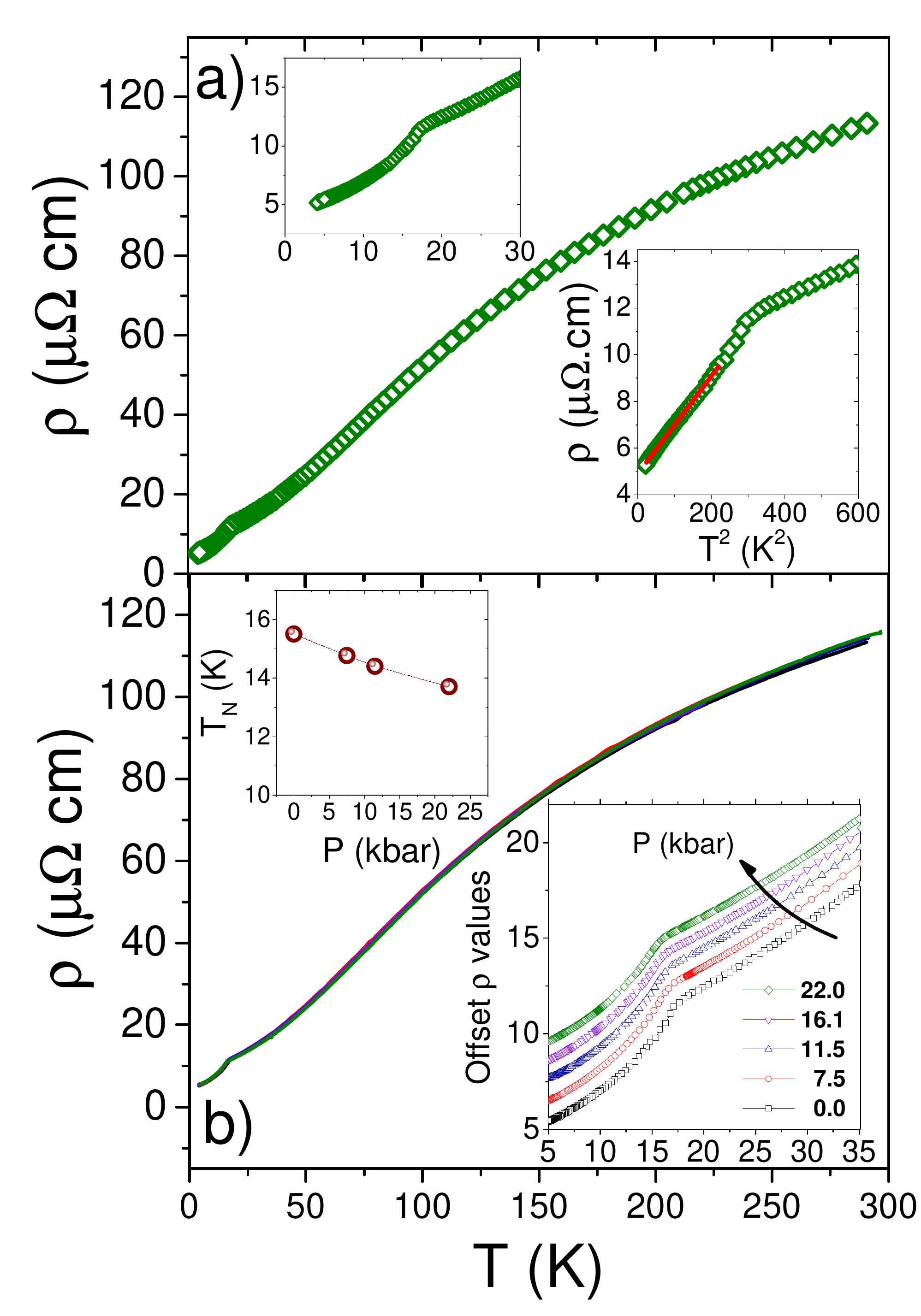}
\end{center}
\vspace{-0.7cm}
\caption{a) Electrical resistivity along the b-axis, $\rho_{b}(T)$, of CaCo$_{5}$As$_{3}$ as a function of temperature. Insets plot the low-temperature resistivity 
versus $T$ and $T^{2}$. b) Pressure dependence of $\rho_{b}(T)$. Left inset shows the pressure dependence of $T_{M}$. Right inset shows the low-$T$ data for various pressures.}
\label{fig:Fig4}
\end{figure}

As in CaFe$_{4}$As$_{3}$ \cite{CaFeAs2}, there is Fermi-liquid regime below 
$15$~K where $\rho(T)=\rho_{0}+AT^{2}$  yields a straight line in a $\rho$ $vs.$ $T^{2}$ plot (bottom inset of Fig.~4a). The extracted values are $\rho_{0} = 4.9$~$\mu \Omega.$cm 
and $A =0.021$~$\mu \Omega.$cm.$K^{-2}$. The corresponding Kadowaki-Woods (KW) ratio is $A/\gamma^{2} = 9a_{0}$, where $a_{0}=10^{-5}$~$\Omega$.cm.mol$_{\mathrm{Co}}^{2}$.K$^{2}$.J$^{-2}$ is 
a universal value
observed in strongly correlated materials. This ratio ($A/\gamma^{2}$) is about $6$ times smaller than that for CaFe$_{4}$As$_{3}$, and not surprisingly, $\chi_{0}$ is about $4$ times and $\gamma$ about $1.4$ times smaller in CaCo$_{5}$As$_{3}$ compared to CaFe$_{4}$As$_{3}$ \cite{CaFeAs2}. Together, these suggest that electronic correlations are present but not as strong in CaCo$_{5}$As$_{3}$ as they are in isostructural CaFe$_{4}$As$_{3}$. 

To investigate whether the magnetic order of CaCo$_{5}$As$_{3}$ can be suppressed to zero temperature, we measured its electrical resistivity as a function of applied pressure. Figure 4b shows that the
overall temperature dependence of $\rho_{b}(T)$ does not change significantly with applied pressure. A closer look at the low-temperature region, however, shows that the magnetic transition temperature
decreases smoothly as a function of pressure (bottom inset of Fig. 4b). The rate of suppression is rather low ($\sim -0.008$~K/GPa) and higher pressures are necessary to fully suppress the magnetic order.

Electronic structure calculations lend additional support to conclusions drawn from experiments and highlight differences between CaCo$_{5}$As$_{3}$ and 
CaFe$_{4}$As$_{3}$. Figure~5 displays the density of states of CaCo$_{5}$As$_{3}$ as a function of energy, for which there is
a finite density of states at the Fermi level contributed by all five of the Co ions. The associated band dispersion, plotted in Fig.~6b, is notably 3-dimensional, and in particular, planar-like sheets parallel to the b-axis present 
in CaFe$_{4}$As$_{3}$ are absent in our material. 
From the solely itinerant electron calculations that ignore Coulomb interactions, the calculated $N(E_{F}) = 14$ states$/(f.u. eV$) is only about $45$~\% of the density of states obtained from heat capacity ($N(E_{F}) = 30$ states$/(f.u. eV)$), and is consistent with a non-trivial role of electronic correlations in CaCo$_{5}$As$_{3}$.

\begin{figure}[!ht]
\begin{center}
\vspace{-0.3cm}
\hspace{-0.75cm}
\includegraphics[width=0.9\columnwidth,keepaspectratio]{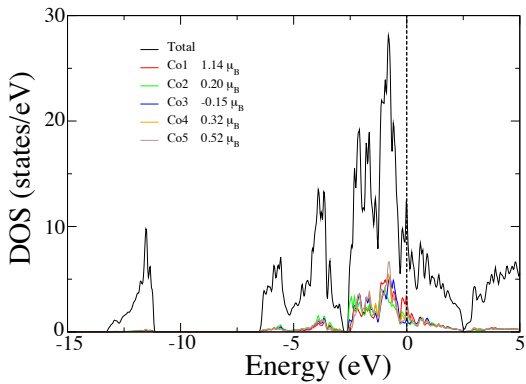}
\end{center}
\vspace{-0.7cm}
\caption{Total density of states (DOS) as a function of energy as well as the partial contribution from each Co site for a non magnetically constrained calculation. The legend provides the moment on each Co site in a spin-polarized calculation.}
\label{fig:Fig6}
\end{figure}

\begin{figure}[!ht]
\begin{center}
\hspace{-0.75cm}
\includegraphics[width=0.62\columnwidth,keepaspectratio]{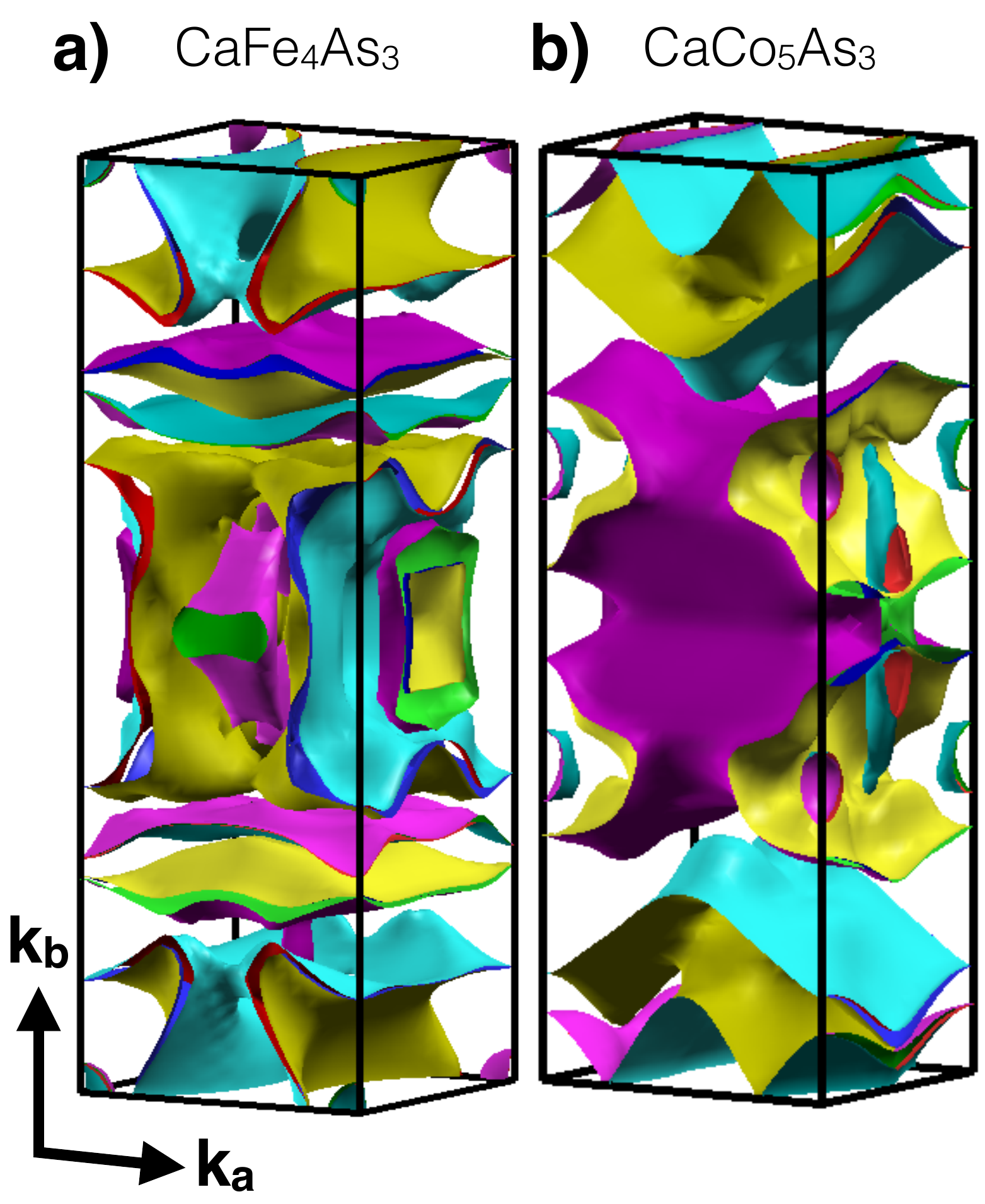}
\end{center}
\vspace{-0.7cm}
\caption{Fermi surface of a) CaFe$_{4}$As$_{3}$ and b) CaCo$_{5}$As$_{3}$.}
\label{fig:Fig7}
\end{figure}

As experimentally measured, the enhanced density of states also suggests a potentially magnetic state. Indeed, spin polarized calculations find a
magnetic ground state. The polarized band structure calculations give distinct magnetic moments for each Co site ($1.14$~$\mu_{B}$ on Co(1), $0.20$~$\mu_{B}$ on 
Co(2), $-0.15$~$\mu_{B}$ on Co(3), $0.32$~$\mu_{B}$ on Co(4), and $0.52$~$\mu_{B}$ on Co(5)), showing clearly that the moment depends on the local environment 
of each Co site. Further, these calculations anticipate ferrimagnetic order, as argued from magnetic susceptibility and magnetization measurements to be the likely ordered structure.

 As mentioned in the Introduction, the Fe moment in the Fe-based superconductors is sensitive to the Fe-As 
hybridization and tends to increase with increasing Fe-As 
distance. This tendency is recreated in CaCo$_{5}$As$_{3}$, as the largest magnetic moment is observed at the site with largest Co-As distance (Co1). The Co-As distances vary from $2.37(6)$~\AA~ for Co2-5 up to $2.66$~\AA~ for Co1. Further, the bond angles Co-As-Co are closer to $90^{o}$ for nearest-neighbors than for next-nearest-neighbors, indicating that nearest neighbors have parallel spins and next-nearest-neighbors 
have antiparallel spins. This effect is in agreement with the observed ferromagnetic component and, as in FeAs, is consistent with the Goodenough-Kanamari rules that arise when local-moment physics is present. 
Thus, our results point to a scenario in which 
 competing exchange interactions in a FM background drive magnetic frustration. This scenario explains the smooth 
pressure-dependent decrease of $T_{N}$  because applied pressure is known to increase the hybridization by decreasing the lattice parameters. It will be interesting to investigate whether higher pressures 
will continuously drive the 
system to a quantum critical point giving rise to a quantum spin liquid state. Another possible scenario is that pressure will relieve frustration, resulting in an increase of $T_{N}$ or a change in the magnetic ground state.

The intricate structure of CaCo$_{5}$As$_{3}$ is further reflected in its Fermi surface, shown in Figure 6b. For comparison we also computed the FS of CaFe$_{4}$As$_{3}$ using the structure determined by Ref. [8]. 
The FS of CaFe$_{4}$As$_{3}$  shown in Fig.~6a is in good agreement with that from Ref.~[8]. The main difference between the Fermi surfaces of CaFe$_{4}$As$_{3}$ 
 and 
CaCo$_{5}$As$_{3}$ is the absence of sheets parallel to the $b$-axis in the Co counterpart. Because 
CaCo$_{5}$As$_{3}$ has 
one additional Co atom filling the `hole' in the crystal structure, the strip-like structure previously seen in CaFe$_{4}$As$_{3}$ is now missing and, as a consequence, no clear sign of `visual' nesting is observed in 
CaCo$_{5}$As$_{3}$. We speculate that the three-dimensionality observed in CaFe$_{4}$As$_{3}$ and CaCo$_{5}$As$_{3}$ is likely responsible for the absence of superconductivity in these materials.

\section{CONCLUSIONS}

In summary,  we have synthesized single crystals of CaCo$_{5}$As$_{3}$, a new transition metal pnictide compound. 
CaCo$_{5}$As$_{3}$ is a metallic magnet ($T_{M}= 16$~K) with a moderate enhancement in electronic 
correlations. 
Its complex magnetic response indicates competing exchange interactions and magnetic frustration. 
Electrical resistivity measurements under pressure to 25~kbar show a smooth decrease in $T_{M}$ to $14$~K. Experiments at higher pressures and neutron diffraction measurements will be valuable to 
fully suppress the magnetic order and to solve the magnetic structure of CaCo$_{5}$As$_{3}$, respectively.

\begin{acknowledgments}
Work at Los Alamos National Laboratory (LANL) was performed under the auspices of the U.S. Department of Energy, Office of Basic Energy Sciences, Division of Materials Science and Engineering. P. F. S. R. acknowledges a Director's Postdoctoral Fellowship through the LANL LDRD program.
\end{acknowledgments}

\bibliography{basename of .bib file}

\end{document}